\documentclass{openjournal}

\usepackage{algorithmic}
\usepackage{bbm}
\usepackage{graphicx}
\usepackage{booktabs}
\usepackage[dvipsnames,table]{xcolor} 
\usepackage{tikz}
\usetikzlibrary{arrows,arrows.meta,automata,positioning,calc}
\makeatletter
\newcommand{\HEADER}[1]{\ALC@it\underline{\textsc{#1}}\begin{ALC@g}}
\newcommand{\ENDHEADER}{\end{ALC@g}}
\newcommand{\algorithmicbreak}{\textbf{break}}
\newcommand{\BREAK}{\STATE \algorithmicbreak}
\makeatother
\usepackage{textgreek}
\usepackage[utf8]{inputenc}
\usepackage[english]{babel} 
\usepackage{amsmath}
\definecolor{linkcolor}{rgb}{0.0,0.3,0.5}
\usepackage{tensind}
\tensordelimiter{?}
\DeclareGraphicsExtensions{.bmp,.png,.jpg,.pdf}
\usepackage{verbatim}
\usepackage[normalem]{ulem}
\usepackage{soul}

\usepackage{hyperref}

\usepackage{orcidlink}
\hypersetup{
unicode,
colorlinks=true,
linkcolor=linkcolor,
citecolor=linkcolor,
filecolor=linkcolor,
urlcolor=linkcolor,
}
\graphicspath{ {./Figures/} }

\begin{document}
\title[\texttt{PolySwyft}]{\texttt{PolySwyft}: sequential simulation-based nested sampling}

\author{K. H. Scheutwinkel,$^{1,2}$}
\email{khs40@cantab.ac.uk}
\author{
W. Handley,$^{1,2}$}
\author{C. Weniger$^{3}$}
\author{E. de Lera Acedo$^{1,2}$}
\affiliation{
$^{1}$Astrophysics Group, Cavendish Laboratory, J. J. Thomson Avenue, Cambridge CB3 0HE, UK\\
$^{2}$Kavli Institute for Cosmology, Madingley Road, Cambridge CB3 0HA, U\\
$^{3}$Gravitation Astroparticle Physics Amsterdam (GRAPPA), University of Amsterdam, Science Park 904, 1098 XH Amsterdam, The Netherlands
}
\begin{abstract}
We present \texttt{PolySwyft}, a novel, non-amortised simulation-based inference framework that unites the strengths of nested sampling (NS) and neural ratio estimation (NRE) to tackle challenging posterior distributions when the likelihood is intractable but a forward simulator is available. By nesting rounds of NRE within the exploration of NS, and employing a principled KL-divergence criterion to adaptively terminate sampling, \texttt{PolySwyft} achieves faster convergence on complex, multimodal targets while rigorously preserving Bayesian validity. On a suite of toy problems with analytically known posteriors of a $\dim(\theta,D)=(5,100)$ multivariate Gaussian and multivariate correlated Gaussian mixture model, we demonstrate that \texttt{PolySwyft} recovers all modes and credible regions with fewer simulator calls than \texttt{swyft}'s TNRE. As a real-world application, we infer cosmological parameters $\dim(\theta,D)=(6,111)$ from CMB power spectra using \texttt{CosmoPower}.  $\texttt{PolySwyft}$ is released as open-source software, offering a flexible toolkit for efficient, accurate inference across the astrophysical sciences and beyond.
\end{abstract}


\maketitle

\section{Introduction}
 Within modern cosmology, it is common \citep{trotta_bayes_2008, trotta_bayesian_2017} to use Bayesian inference methods for analysing datasets to probe the cosmological evolution of the universe. Bayesian algorithms such as Markov Chain Monte Carlo (MCMC) methods \citep{metropolis_equation_1953, mackay_information_2003} are used in practice \citep{dunkley_fast_2005, christensen_bayesian_2001, knox_age_2001, lewis_cosmological_2002, verde_first-year_2003, tegmark_cosmological_2004} to estimate cosmological parameters for a given hypothesis while many other researchers \citep{scheutwinkel_bayesian_2022, bevins_comprehensive_2022, anstey_general_2021, shen_quantifying_2021,handley_quantifying_2019, hergt_bayesian_2021} now use nested sampling \citep{skilling_nested_2006, sivia_data_2006} instead to conduct model comparison and parameter estimation simultaneously. As the broader (cosmological) scientific field now requires and has access to more advanced data analysis methods \citep{ntampaka_role_2019}, limitations within Bayesian inference have started to emerge. 

One of these limitations is the intractability of the likelihood function for state-of-the-art cosmological models; hence, simulation-based inference (SBI), also known as likelihood-free inference (LFI) methods, was developed \citep{marin_approximate_2011, gutmann_bayesian_2016,cranmer_frontier_2020, lueckmann_benchmarking_2021}, forming a new statistical paradigm within Bayesian inference. As the names of LFI and SBI suggest, the statistical modelling process does not demand an explicit likelihood expression, thus it is ``free'' of statistical assumptions and a simulator-driven approach is used instead. 

Generally, in the sciences, one can access a theoretical simulator that takes in a parameter vector $\theta$ and forward models a dataset $D$. SBI methods have been used within cosmology ~\citep{alsing_fast_2019, jeffrey_likelihood-free_2020, cole_fast_2021, zhao_implicit_2022, zhao_simulation-based_2022, lin_simulation-based_2023, von_wietersheim-kramsta_kids-sbi_2024, jeffrey_dark_2024, saxena_constraining_2023, saxena_simulation-based_2024, anaumontel_estimating_2023, bhardwaj_sequential_2023, gagnon-hartman_debiasing_2023, karchev_sicret_2023, alvey_albatross_2023} with architectures such as \texttt{swyft} \citep{miller_simulation-efficient_2020, miller_truncated_2021} that uses neural networks as binary classifiers to approximate likelihood ratios \citep{cranmer_approximating_2016, thomas_likelihood-free_2020}. Moreover, sequential learning methods have been developed \citep{papamakarios_sequential_2019, lueckmann_flexible_2017, wiqvist_sequential_2021, dirmeier_simulation-based_2025, haggstrom_fast_2024, rubio_transport_2023, sharrock_sequential_2024, deistler_truncated_2022}, where one iteratively retrains a neural network based on the previous network predictions as a form of active learning to accelerate retraining efforts to find a better estimator. \texttt{swyft} implements such a sequential learning method for retraining a better NRE by iteratively truncating the prior driven by an observation $D_\mathrm{obs}$, making it non-amortised. Limitations with this method arise, as constructing efficient truncation schemes for highly multimodal problems remains challenging.

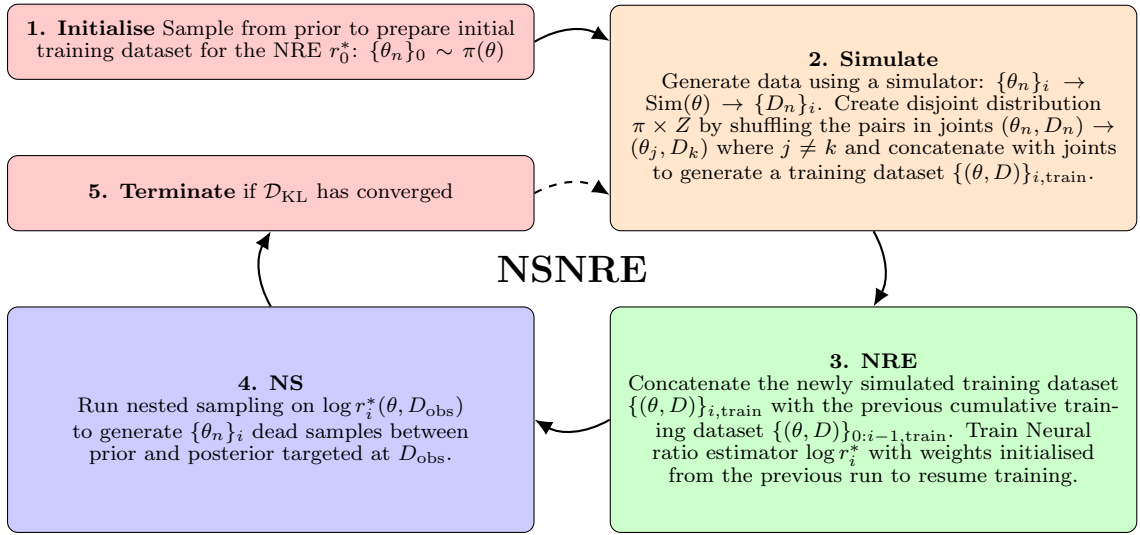
\begin{figure*}
    \centering
    \begin{tikzpicture}[
            node distance=1cm,
            >=stealth, auto,
            every state/.style={
                rectangle, rounded corners, minimum width=2em,
                text width=6.8cm, align=center
            }
        ]
          \node[state, fill=blue!20, minimum height=3cm] (q14) { 
            \textbf{4. NS}\\
            Run nested sampling on $\log r^*_{i}(\theta, D_\mathrm{obs})$ \\
            to generate $\{\theta_n\}_i$ dead samples between prior and posterior targeted at $D_\mathrm{obs}$.

        };
        \node[state, fill=red!20, minimum height=1cm] (q121) [above=of q14] { 
            \textbf{5. Terminate} if $\mathcal{D}_\mathrm{KL}$ has converged
        };
        \node[state, fill=red!20, minimum height=1cm] (q12) [above=of q121] {
            \textbf{1. Initialise} Sample from prior to prepare initial training dataset for the NRE $r^*_0$: $\{\theta_n\}_0 \sim\pi(\theta)$ \\
        };
        \node[state, fill=green!20, minimum height=3cm] (q34) [right=of q14] { 
            \textbf{3. NRE}\\
            Concatenate the newly simulated training dataset $\{ (\theta,D)\}_{i, \mathrm{train}}$ with the previous cumulative training dataset $\{ (\theta,D)\}_{0:i-1, \mathrm{train}}$.
            Train Neural ratio estimator $\log r^*_i$
            with weights initialised from the previous run to resume training. };
        \node[state, fill=orange!20, minimum height=3cm] (q23) [above=of q34] {
            \textbf{2. Simulate}\\
            Generate data using a simulator: $\{\theta_n\}_i \to \mathrm{Sim(\theta)} \to  \{D_n\}_i$. Create disjoint distribution $\pi \times Z$ by shuffling the pairs in joints $(\theta_n, D_n) \to (\theta_j, D_k)$ where $j \neq k$ and concatenate with joints to generate a training dataset $\{ (\theta,D)\}_{i, \mathrm{train}}$.
        };
        \begin{scope}[bend left]%
            \path[thick,-{Latex[width=2mm]}]   (q14.north) edge node {} (q121.south)
            (q121.east) edge[dashed] node {}($(q23.south west)!0.16!(q23.north west)$) 
            (q23.south) edge node {} (q34.north)
            (q34.west) edge node {} (q14.east)
            (q12.east) edge node {}($(q23.south west)!0.84!(q23.north west)$) ;
        \end{scope}

        \node[align=center] (e) at (barycentric cs:q121=0.5,q12=0.5,q23=1,q34=1,q14=1) {\Large\textbf{NSNRE}};
    \end{tikzpicture}
    \caption{Nested Sampling Neural Ratio Estimation (NSNRE) meta-algorithm cycle. There are 5 distinct phases: 1. Sample from the prior for the initial training dataset 2. Use a forward simulator to sample joint $(\theta,D)$ 3. (Re-)Train an NRE on joint $\mathcal{J}$ and disjoint $\pi \times Z$ dataset 4. Use an NS on NRE to generate new samples $\theta$ around observation $D_{\mathrm{obs}}$. 5. Terminate if the KL-divergence criterion is fulfilled; otherwise, continue with step 2 (dashed arrow).}
    \label{fig:NSNRE_meta_algo}
\end{figure*}

To address this issue, we propose \texttt{PolySwyft}, which does not require a prior truncation scheme and explores the full ``likelihood-to-evidence'' space instead, which can be multimodal or complex-shaped. This marginal-free approach is an alternative to \texttt{swyft}'s marginal TNRE approach to recover higher-dimensional posterior distributions. The meta-algorithm idea is shown in Figure~\ref{fig:NSNRE_meta_algo}.
In section~\ref{sec:Bayes}, we describe the theory of the algorithm components of \texttt{PolySwyft}. In section~\ref{sec:PolySwyft}, we present \texttt{PolySwyft} and assess it on toy problems in section~\ref{sec:Toy}. In section~\ref{sec:future_improv}, we discuss future improvements on \texttt{PolySwyft}. Finally, we conclude in section~\ref{sec:Conclusion}. 

\section{Bayesian Inference}
\label{sec:Bayes}
With Bayesian inference, one assigns a prior belief $\pi \equiv p(\theta|M)$ of the hypothesis $M$ into a probabilistic distribution and incorporates this randomness into the probabilistic modelling of the dataset $D$ known as the likelihood $\mathcal{L} \equiv p(D|\theta, M)$. Through Bayes Theorem, these quantities are dependent in the following way: \begin{equation}
\mathcal{P} \times Z = \mathcal{L} \times \pi = \mathcal{J},
\end{equation}
where $\mathcal{J} \equiv p(\theta,D |M)$ is the joint ``probability of everything'' distribution, $\mathcal{P} \equiv p(\theta | D, M)$ is the posterior belief of the hypothesis after the experiment is conducted and $Z \equiv p(D|M) $ is the marginalised likelihood over the prior space. We also define the disjoint distribution $\pi \times Z \equiv p(\theta|M)p(D|M)$. The posterior is a probabilistic distribution of the parameter space $\theta$, commonly used as the quantity in the parameter estimation part of Bayesian inference. Whereas $Z$, also known as the Bayesian evidence, is a metric to assess how well the hypothesis describes the dataset and is used for model comparison within Bayesian inference.

\subsection{Nested Sampling}
Usually, it is challenging to analytically derive the posterior distribution for a given model of an underlying physical system. Hence, algorithmic methods such as MCMC methods \citep{mackay_information_2003} are executed to estimate these distributions through a sampling-based approach. However, MCMC methods do not solve for the marginalised likelihood, as the underlying sampling mechanism sets the Bayesian evidence as a constant factor for a given observation $D_\mathrm{obs}$. Thus, nested sampling \citep{skilling_nested_2006, sivia_data_2006} as an alternative method is utilised. The advantage of the nested sampler is due to its sampling-based approach of solving for the marginalised likelihood; the accumulated samples during the algorithm execution are posterior weighted. Therefore, a nested sampler provides a ``free'' set of posterior samples next to the marginalised likelihood estimate. The marginalised likelihood is defined as:
\begin{equation}
\label{eqn:NS_int}
    Z \equiv p(D|M) = \int_{\theta \in \Theta} p(D|\theta,M) p(\theta|M) d\theta = \int_0^1\mathcal{L} dX,
\end{equation}
where $dX = p(\theta) d\theta$ is the prior volume fraction element. To numerically estimate this integral, one constructs prior volume shells $X$ that contains a minimum hard likelihood constraint $\mathcal{L}^*$. This likelihood constraint evolves while the prior volume fraction is shrinking as the algorithm processes:
\begin{equation}
    X(\mathcal{L}^*) = \int_{\mathcal{L} > \mathcal{L}^*} p(\theta |M ) d\theta.
\end{equation}
A schematic nested sampling algorithm for solving the integral in eq.~\eqref{eqn:NS_int} can be seen in Algorithm~\ref{algo:NS}.

\begin{figure}[htb!]
  \caption{Nested sampling algorithm with a generic Sampler.}
    \label{algo:NS}
  \begin{algorithmic}
  \HEADER{Input}
    \STATE $\pi(\theta)$, $\mathcal{L}(\theta|D_\mathrm{obs})$, $n_{\mathrm{live}}$, $\epsilon_{\mathrm{NS}}$; 
    \ENDHEADER
  \HEADER{Initialize}
  \STATE $Z_{0} = 0$, $X_0 = 1$, $Z_{\mathrm{increase,0}} = 10^{30}, j = 0$
  \ENDHEADER
  \STATE  Draw $ \{ \theta_i \}_{i=1}^{n_{\mathrm{live}}} \sim \pi(\theta)$; 
  \STATE Evaluate $\{ \mathcal{L}_i\}_{i=1}^{n_{\mathrm{live}}} \leftarrow \mathcal{L}(\theta_i|D_\mathrm{obs})$;
  \WHILE{$Z_{\mathrm{increase,j}}  > \epsilon_{\mathrm{NS}}$}
  \STATE $\mathcal{L}_{\min,j} \leftarrow \min \{ \mathcal{L}_i\}$;

  \STATE 
      Draw $\theta^*_j \sim  \mathrm{Sampler}(\theta)$ s.t. $\mathcal{L}(\theta^*_j) > \mathcal{L}_{\min,j}$ and $\theta^*_j \in \pi(\theta)$
    \STATE Replace $\theta_{\min, j}$ with $\theta_j^*$ in current live points set $\{\theta_i \}_{i=1,j}^{n_{\mathrm{live}}}$ 
    \STATE Draw contraction $t_j \sim p(t|n_{\mathrm{live}})$;
    \STATE $X_{j+1} \leftarrow  t_j \cdot X_j$;
    \STATE $\Delta X_j \leftarrow  X_j - X_{j+1}$;
    \STATE $Z_{j+1} \leftarrow  Z_j + \mathcal{L}_{\min,j} \Delta X_j$; 
    \STATE $Z_{\mathrm{increase,j+1}} \leftarrow  \frac{\mathcal{L}_{\max,j} \Delta X_j}{Z_{j+1}}$;
    \STATE $j \leftarrow j + 1$;
  \ENDWHILE
  \STATE $Z_{\mathrm{total}} \leftarrow  Z_{\mathrm{total}, j} + \frac{1}{n_{\mathrm{live}}} \sum_{i=1}^{n_{\mathrm{live}}} \mathcal{L}_i \Delta X_i $;
    \RETURN $Z_{\mathrm{total}}$, $\{(\theta, \mathcal{L}) \}$ 
\end{algorithmic}
\end{figure}

We chose \texttt{PolyChord} \citep{handley_PolyChord_2015, handley_PolyChord_2015-1} as our nested sampler, as its slice sampling-based mechanism \citep{neal_slice_2000} and K-nearest neighbour clustering functionality for multimodality detection proves to be an efficient and scalable solution for higher dimensional parameter spaces. We note that alternative nested samplers exist which are presented in the survey by \cite{ashton_nested_2022, buchner_nested_2023}.

\subsection{Neural ratio estimation}
Neural Ratio Estimation (NRE) uses a Neural Network (NN) to estimate the ratio between two quantities. More concretely in the context of SBI, with a NN and its parameters $\phi$, one estimates the likelihood-to-evidence ratio $r_\phi (\theta, D)$:
\begin{equation}
    r_\phi(\theta, D) \approx \frac{p(D|\theta)} {p(D)} = \frac{p(\theta,D)}{ p(\theta) p(D)}  \equiv \frac{\mathcal{J}}{\pi \times Z} =  \frac{\mathcal{L}}{Z} = \frac{\mathcal{P}}{\pi},
\end{equation}
which is ratio of the joint distribution $\mathcal{J}$ over the disjoint distribution $\pi \times Z$. One can cast this ratio into the following classification problem by introducing categorical ``labels'' $M$:
\begin{equation}
   p(\theta,D|M)= \begin{cases}
    p(\theta,D), \ \text{if } M =M_{\mathcal{J}}  \\
    p(\theta)p(D), \ \text{if } M = M_{\pi Z}
\end{cases},
\end{equation} where $M_{\mathcal{J}}: (\theta, D) \sim \mathcal{J}$ and $M_{\pi Z}: (\theta,D) \sim \pi \times Z$ state whether a $(\theta,D)$ pair was generated from the joint $\mathcal{J}$ or disjoint distribution $\pi \times Z$; with equiprobable model probabilities $p(M_{\mathcal{J}}) = p(M_{\pi Z})= \frac{1}{2}$. With these definitions, one can rewrite the ratio to:
\begin{equation}
\begin{split}
   r_\phi(\theta, D) \approx  \frac{p(\theta,D)}{p(\theta) p(D)} = \frac{p(\theta,D |M_\mathcal{J}) p(M_\mathcal{J})}{p(\theta,D |M_{\pi Z}) p(M_{\pi Z})} \\= \frac{p(M_\mathcal{J} | \theta,D )}{p(M_{\pi Z} | \theta,D )}  = \frac{p(M_\mathcal{J} | \theta,D )}{1-p(M_{\mathcal{J}} | \theta,D )},
\end{split} 
\end{equation} where one defines a classifier $p(M_{\mathcal{J}}|\theta,D) = 1-p(M_{\pi Z} |\theta, D)$.
Thus, one can estimate the ratio by training a binary classifier for $p(M_{\mathcal{J}}| \theta,D)$, also known as the likelihood-ratio trick \citep{cranmer_approximating_2016}. One trains this classifier by minimising a binary cross-entropy loss function:
\begin{equation}
\begin{split}
   &  \mathbb E_{p(M,\theta,D)}  \left[ -p(M|\theta, D) \right] \\ & =  
     -\mathbb E_{p(\theta,D)} \left[  p(M_\mathcal{J}|\theta,D)\right]  - \mathbb E_{p(\theta) p(D)} \left[ p(M_{\pi Z}|\theta,D) \right].
\end{split}
\end{equation}
To generate the dataset for this binary classification problem, one needs a simulator for this approach. This (probabilistic) simulator acts as a surrogate sampler of the intractable likelihood function $p(\cdot|\theta_i)$ with $\theta_i \sim \pi(\cdot)$, therefore forming the samples from the joint distribution. The disjoints are acquired by shuffling the set of joints draws $\{(\theta_i, D_i)\}$ with each other. Usually for a physical problem, a simulator can be accessed that takes a pair of input parameters $\theta_i$ and generates the corresponding dataset $D_i$.

An implementation of this NRE approach is \texttt{swyft} \citep{miller_truncated_2021, cole_fast_2021} that extends this methodology to Truncated Marginal NRE (TMNRE) by training multiple classifiers of marginals of parameter pairs instead of the full joint space and sequentially truncates the prior regions to estimate the posterior distribution for a given observation $D_\mathrm{obs}$.

\section{\texttt{PolySwyft}}
\label{sec:PolySwyft}
TMNRE is an algorithm that can accurately recover posterior estimates of high-dimensional problems and has been used in practice in 21-cm cosmology ~\citep{saxena_constraining_2023, saxena_simulation-based_2024}, strong lensing ~\citep{anaumontel_estimating_2023}, gravitational wave detection ~\citep{bhardwaj_sequential_2023}, and other cosmological problems ~\citep{gagnon-hartman_debiasing_2023, karchev_sicret_2023, alvey_albatross_2023}. However, there are limitations in the truncation scheme of this methodology. \texttt{swyft}'s TNRE truncates regions in the prior that have less weight to the observed data $D_\mathrm{obs}$ and iteratively constructs shrinking regions in the shape of rectangles through the cumulative distribution function (CDF) and uses its inverse to sample within the truncated bounds. This approach has proven to be useful for problems that have posterior distributions that are unimodal and of simple shape. However, limitations arise if multimodality is expected or for posteriors with complex shapes. This rectangular truncation scheme becomes increasingly difficult for these problems, and sampling efficiency becomes exponentially worse for higher dimensions. We propose a new truncation scheme with \texttt{PolySwyft} to address these issues. \texttt{PolySwyft} is a combination of \texttt{PolyChord} and \texttt{swyft} that sequentially executes these two algorithms - nested sampling and neural ratio estimation - until a termination criterion is fulfilled or a pre-determined number of rounds have passed, as presented in Figure~\ref{fig:NSNRE_meta_algo}. 

\subsection{Dead measure}
\label{subsec:dead}
\begin{figure*}
    \centering
\includegraphics{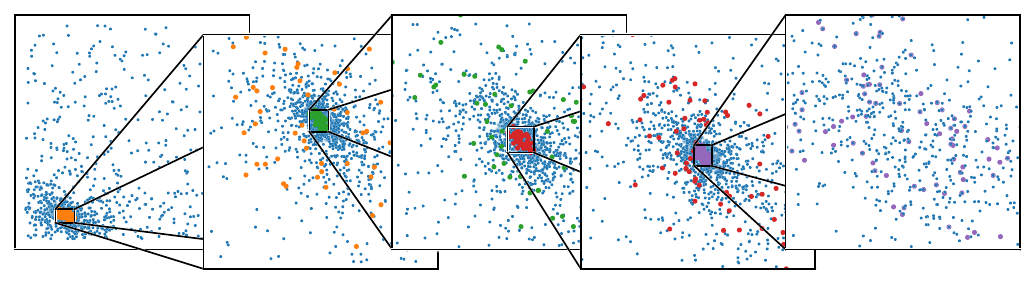}
    \caption{A typical dead points distribution for a parameter pair where one recursively zooms into the exponentially dense regions of dead points. The dead points have constant density in $\log X$, while the live points (larger coloured dots) have uniform density until termination. The plots were generated with code provided by \cite{hu_aeons_2023}.}
    \label{fig:deadmeasure}
\end{figure*}

\texttt{PolySwyft} uses nested sampling as a scheme to explore regions driven by the observation $D_\mathrm{obs}$ within the likelihood-to-evidence space of the \texttt{swyft} NRE $r_\phi (\theta,D)$ and sequentially generate samples $\theta$ that satisfy a minimum ratio criterion $r(\theta,D_{\mathrm{obs}}) > r_{\mathrm{min}}$. These samples, known as dead points, have a distribution that initially uniformly fills the whole prior space but exponentially populates samples in regions of higher ratio estimates by constructing contracting prior volume shells $X$. This distribution is the dead measure $\pi^*(\theta)$ that was generated from an initial prior distribution $\pi(\theta)$. An example of a typical dead measure generated through a nested sampler is shown in Figure~\ref{fig:deadmeasure}, where generally $\pi(\theta) \neq \pi^*(\theta)$.

Hence, the nested sampling process is considered a ``truncation mechanism'' in high-ratio regions that substitutes the prior truncation scheme in T(M)NRE. Nested sampling exponentially populates dead points around the posterior maximum, thus acting as a surrogate ``truncation mechanism'' into the regions driven by $D_{\mathrm{
obs}}$. Through this approach, the detection and population of multimodal regions depend on the nested sampling algorithm design, which is set as $\texttt{PolyChord}$ here. 

This intrinsic mechanism of contracting shells generates more samples that are further concentrated around the posterior peak if one continuously samples via nested sampling. However, as one is generally interested in estimating the posterior distribution rather than approximating the posterior peak, generating more samples close to the peak can be inefficient for subsequent neural network training. We terminate the nested sampling algorithm as in the usual setting i.e. the current evidence estimate that is contained by the set of live points is less than a fraction $\epsilon_{\mathrm{NS}}$ of the accumulated evidence so far, elaborated in section~\ref{subsec:init}. 

Once $\texttt{PolyChord}$ has terminated and generated a new set of dead points $\{\theta\}$, these dead points are fed into the simulator to create a new training dataset that is concatenated with the previous training dataset. This cumulative dataset forms the new training dataset for the next ratio estimator $r_{\phi,i+1} (\theta,D)$. This dataset concatenation is a form of active learning to enhance the retraining procedure with more relevant samples i.e.\ forming a sequential method, while simultaneously mitigating the effects of catastrophic forgetting \citep{mccloskey_catastrophic_1989, ratcliff_connectionist_1990} by keeping the initial training dataset drawn from the prior. We use standard practitioners' methodologies ~\citep{Goodfellow-et-al-2016, geron_hands_2019} of early stopping with patience on the validation loss to terminate neural network (re-)training.

We highlight here that the training dataset of $\texttt{PolySwyft}$ has a different distribution than T(M)NRE as $\texttt{PolySwyft}$'s NRE $r_{\phi,i+1}(\theta, D_{\mathrm{obs}})$ is trained on a concatenation of dead measures $\tilde\pi^*_i(\theta)=\omega_{\mathrm{\pi}} \pi(\theta) +\sum_{i=0}^{N_{\mathrm{iter}}} \omega_i\pi_{i}^*(\theta)$ while \texttt{swyft} is sequentially truncating the original prior $\pi_{\Gamma,i}(\theta) = \pi(\theta) \times \prod_{i=0}^{N_{\mathrm{iter}}}V^{-1}_i \Gamma_i(\theta)$.

\subsection{The algorithm}
The algorithm that $\texttt{PolySwyft}$ executes consists of a cyclic execution of an NRE method and nested sampling. For $\texttt{PolySwyft}$, an implementation of this method, we have the following procedure presented in Algorithm~\ref{algo:polyswyft}.

\begin{figure}[htb!]
  \caption{\texttt{PolySwyft} algorithm}
    \label{algo:polyswyft}
  \begin{algorithmic}
  \HEADER{Inputs}
  \STATE $\mathrm{Sim}, \mathrm{NRE}, D_{\mathrm{obs}},  N^{(0)}, N_{\mathrm{iter}} , C_{\mathrm{comp.}}$ 
  \ENDHEADER
  \HEADER{Initialise}
  \STATE $(\theta,D)^{0}  = \{D_n \sim 
  \mathrm{Sim}(\theta_n), \theta_n \sim \pi(\theta)\}_{n=1}^{N^{(0)}}$
  \ENDHEADER
  \FOR{$i=0$ in $N_{\mathrm{iter}}$} 
    \HEADER{\texttt{Swyft (or any NRE)}}
      \STATE $r^*_i \leftarrow \mathrm{NRE}((\theta,D)^{(i)})$ 
    \ENDHEADER
    \HEADER{\texttt{PolyChord (or any NS)}} 
      \STATE $\{\theta_{n}\}_{n=1}^{N^{(i)}} \leftarrow \mathrm{NS}(r^*_i,D_{\mathrm{obs}}, \pi(\theta))$ 
    \ENDHEADER
      \STATE $\mathrm{KL}^{(i)}_{\mathrm{comp.}} \leftarrow \mathrm{KL}(\mathcal{P}_{i}||\pi)$
      \IF{$i > 0$}
      \STATE $\mathrm{KL}^{(i)}_{\mathrm{rel.}} \leftarrow \mathrm{KL}(\mathcal{P}_{i}||\mathcal{P}_{i-1})$
      \IF{$\mathrm{KL}^{(i)}_{\mathrm{rel.}} \approx 0$ \AND $\mathrm{KL}^{(i)}_{\mathrm{comp.}} > C_{\mathrm{comp.}}$}
      \BREAK
      \ENDIF
      \ENDIF
      \STATE $i \leftarrow i+1$
    \STATE $(\theta,D)^{(i)} \leftarrow (\theta,D)^{(i-1)}  \cup \{D_n \sim 
  \mathrm{Sim}(\theta_n), \theta_n \}_{n=1}^{N^{(i)}}$
  \ENDFOR
\end{algorithmic}
\end{figure}

\subsection{Initialisation}
\label{subsec:init}
$\texttt{PolySwyft}$ has various initialisation settings inherited from \texttt{PolyChord} and \texttt{swyft} that can be fine-tuned. The most relevant parameters with the largest influence on the inference results are the parameters associated with the generation of dead points that are used for retraining, for instance, $\texttt{precision\_criterion}$, $n_{\mathrm{live}}$ and $n_{\mathrm{lives}}$. The parameter $\texttt{precision\_criterion}$ sets the termination criterion of the nested sampling run. Here, the criterion is defined as the fraction $\epsilon_{\mathrm{NS}}$ of the contained estimated evidence of the current live points to the current accumulated evidence. In its standard settings, if $\epsilon_{\mathrm{NS}} < 0.001$, terminate. $n_{\mathrm{live}}$ sets the constant number of live points to maintain within a hard ratio iso-contour, while $n_{\mathrm{lives}}$ adjusts the dynamic number of live points at a given ratio contour.

\subsection{Termination criterion}

For the termination criterion, we can define the KL-divergence \citep{kullback_information_1951} to compare relative changes of the posterior estimates at iteration $i$ and $i-1$ as $\texttt{PolySwyft}$ progresses:

\begin{equation}
\label{eqn:KL}
    \mathrm{KL}(\mathcal{P}_i||\mathcal{P}_{i-1}) = \int  \log \left(\frac{p_i(\theta|D_\mathrm{obs})}{p_{i-1}(\theta |D_\mathrm{obs})} \right) p_i(\theta|D_\mathrm{obs}) d\theta.  
\end{equation}

To derive the posterior distributions $p_{i}(\theta|D_\mathrm{obs})$, one needs to introduce a correction term $Z_{\mathrm{NS}}$ that is computed through a nested sampling run, here with $\texttt{PolyChord}$, of the NRE $r^*_i$: 
\begin{equation}
\label{eqn:ZNS}
    Z_{\mathrm{NS}} = \int r^* \pi d\theta = \int \frac{L}{Z_{\mathrm{dead}}} \pi d \theta   = \frac{Z}{Z_\mathrm{dead}},
\end{equation}
where $Z_\mathrm{dead} = \int L \pi_{\mathrm{dead}} d\theta$ is the normalisation constant that is estimated when an NRE is trained on a dead measure distribution $\pi_{\mathrm{dead}} \neq \pi$.
Using eq. ~\eqref{eqn:ZNS} we can expand our ratio estimator expression to:
\begin{equation}
\label{eqn:r_to_ZNS}
        r^* \frac{Z_\mathrm{dead}}{Z} =  \frac{r^*}{Z_\mathrm{NS}} = \frac{L}{Z} = r,
\end{equation} to find the prior corrected ratio estimate $r$.

The posterior $\mathcal{P}$ can then be derived through Bayes theorem:
\begin{equation}
\label{eqn:PosterorEstimate}
    \mathcal{P} = \frac{L}{Z} \pi = \frac{r^*}{Z_\mathrm{NS}} \pi.
\end{equation}

Finally, with eq.~\eqref{eqn:PosterorEstimate} one can derive for the KL expression in eq.~\eqref{eqn:KL}:
\begin{equation}
    \begin{split}
    \label{eqn:NREcomp_corr}
    \mathrm{KL}(\mathcal{P}_i || \mathcal{P}_{i-1}) &  \approx \sum_{n=1}^{N} w_n [\log r_i(\theta_n,D_\mathrm{obs})  - \log Z_\mathrm{NS,i}  \\ & - \log  r_{i-1}(\theta_n, D_\mathrm{obs})  +   \log Z_\mathrm{NS,i-1}  ]
        \end{split}
    \end{equation}
where $\sum_{n=1}^{N} w_n = 1$ and $\theta_{n} \sim p_i(\theta|D_\mathrm{obs})$.

This definition defines the termination criterion as when $\mathrm{KL}(\mathcal{P}_i||\mathcal{P}_{i-1}) \approx 0$. However, in practice, $\texttt{PolySwyft}$ might reach $\mathrm{KL} \approx 0$ when posterior estimations are still inaccurate, i.e. not much compression relative to the prior has happened. This compression largely depends on the underlying problem and $\texttt{PolySwyft}$ initialisations that generate a new training dataset and steer the retraining process through active learning. We, therefore, need to assess the current compression $\mathrm{KL}(\mathcal{P}_i||\pi)$ in addition to the relative change of posterior estimates.

We can derive the compression of the prior to the posterior as:
\begin{equation}
\begin{split}
         \mathrm{KL}(\mathcal{P}||\pi)  & = \int \log \frac{\mathcal{P}}{\pi} \mathcal{P} d\theta \\ & =  \int  \log r(\theta,D_\mathrm{obs}) p(\theta|D_\mathrm{obs}) d\theta  = \mathbb{E}_{p(\theta|D_\mathrm{obs})}\left[ \log r \right]
\end{split}
\end{equation}

We can estimate this quantity through eq.~\eqref{eqn:r_to_ZNS}:
\begin{equation}
     \mathrm{KL}(\mathcal{P}_i||\pi)  \approx \sum_{n=1}^{N}  w_n \log r_i^*(\theta_n, D_\mathrm{obs}) - \log Z_{\mathrm{NS},i},
\end{equation}
where $\sum_{n=1}^{N} w_n = 1$ and $\theta_{n} \sim p_i(\theta|D_\mathrm{obs})$.

Hence, $\texttt{PolySwyft}$ should be terminated if $\mathrm{KL}(\mathcal{P}_i || \mathcal{P}_{i-1}) \approx 0$ with a satisfactory compression $\mathrm{KL}(\mathcal{P}_i||\pi) > C_\mathrm{comp.}$. The user observes these quantities for each iteration of the algorithm and must make informed decisions about termination.

\subsection{Comparison with Ground truth}
For the multivariate Gaussian model and the Gaussian mixture model toy problems in Section~\ref{sec:Toy}, we can derive the analytical ground truth of the posterior $\mathcal{P}_{\mathrm{true}}$. Thus, we can compare \texttt{PolySwyft}'s performance with the ground truth as the algorithm processes with the sequential training. Using our two criteria for termination, we can derive the ground truth values  $\mathrm{KL}(\mathcal{P}_{\mathrm{true}} || \mathcal{P}_{i})$ and $\mathrm{KL}(\mathcal{P}_{\mathrm{true}}||\pi)$:
\begin{equation}
\begin{split}
\label{eqn:KL_GroundTruth_relative}
    \mathrm{KL}(\mathcal{P}_{\mathrm{true}}|| \mathcal{P}_i) & \approx \sum_n w_n [\log \mathcal{P}_{\mathrm{true}}(\theta_n)
    - \log r^*(\theta_n,D_\mathrm{obs})  \\ &  - \log \pi(\theta_n) + \log Z_\mathrm{NS} ],
\end{split}
\end{equation}
and 
\begin{equation}
\label{eqn:KL_GroundTruth_prior}
    \mathrm{KL}(\mathcal{P}_{\mathrm{true}}|| \pi) \approx \sum_n w_n \left(\log \mathcal{P}_{\mathrm{true}}(\theta_n)  - \log \pi(\theta_n) \right),
\end{equation} for $\theta_n \sim \mathcal{P}_{\mathrm{true}}$ and $\sum_n w_n = 1$

\subsection{Comparison with TNRE posterior estimates}
We compare \texttt{PolySwyft}'s likelihood-to-evidence ratio ``truncation scheme'' with the prior truncation scheme of \texttt{swyft}. As \texttt{PolySwyft} explores the full joint parameter space rather than its marginal parameter space, we compare \texttt{swyft}'s truncated NRE functionality (TNRE) with \texttt{PolySwyft} in section~\ref{sec:Toy}. We note that the TNRE by $\texttt{swyft}$ is not the recommended setting for the $\texttt{swyft}$ package, as it has been shown \citep{miller_truncated_2021} that training a marginal-free NRE remains challenging for high-dimensional posteriors. Hence, the TMNRE of the $\texttt{swyft}$ package was developed to address the limitations of TNRE through marginalisation and is therefore its preferred setting. We intentionally explore a marginal-free approach with \texttt{PolySwyft} to enhance the capabilities of the (T)NRE of \texttt{swyft} via nested sampling. This allows us to capture non-linear relationships between posterior parameters as an alternative to the TMNRE extension.

\subsection{Comparison with $\texttt{PolyChord}$}
We compare \texttt{PolySwyft} with $\texttt{PolyChord}$ as we have a likelihood expression available for our examples studied. We use a standard initialisation scheme for $\texttt{PolyChord}$ to provide a vanilla baseline comparison for the posterior estimates. More concretely, we have a fixed number of live points that linearly scale $n_{\mathrm{live}}= c \times n_{\dim}$ by a constant factor $c$ depending on the problem, e.g.\ dimensionality and/or multimodality. We terminate the nested sampling with the standard evidence-based termination criterion of $\epsilon_{\mathrm{NS}} < 0.001$.  Hence, the initialisation scheme is identical to the nested sampling phase of \texttt{PolySwyft} (mentioned in section~\ref{subsec:init}) for each example studied.

To compare the sample efficiency between $\texttt{PolyChord}$ and $\texttt{PolySwyft}$, we use the performance bottleneck of each algorithm as the efficiency metric. For $\texttt{PolyChord}$, the number of likelihood calls that was needed to generate the posterior distribution is a more informative metric of nested sampling than the number of dead points, as often times the likelihood function is the expensive component of a nested sampler. For $\texttt{PolySwyft}$, we can identify the bottleneck as the number of simulator calls, which is the number of samples used to recover the posterior distribution. Although $\texttt{PolySwyft}$ also uses nested sampling, the likelihood calls substituted by an NRE are computationally cheaper than the simulator or a ``real'' likelihood function, therefore negligible for the efficiency metric.

\subsection{Optimizations}
\label{subsec:optimis}
\texttt{PolySwyft} has various optimisations for its retraining meta-algorithm cycle that centre around optimising the dead measure or the network (re-)training scheme for subsequent rounds, thus increasing sample efficiency. In practice, these can be considered fine-tuning methods that optimise the statistical power introduced by the current dead points relative to the accumulated training dataset so far. We note that these optimisation tools should be considered if the standard settings do not yield satisfactory posterior estimates. For instance, starting with a very large training dataset is generally not recommended, as the subsequent nested sampling process will need to generate a significantly large amount of dead points to meaningfully steer the neural network training towards the relevant posterior regions. Similarly, if the problem has a wider prior, one should start with a reasonable-sized training dataset of $\mathcal{O}(10^4 - 10^5)$ so the initial retraining efforts are not wasted for finding a reasonable first starting point.

\subsubsection{Data compression network}
\label{subsec:data_comp}
For problems where the dimensionality of the dataset is high, one can compress the dataset through a neural compression network $C_\phi (x)$ to a lower manifold $s$ that represents the summary statistics. Therefore, the NRE tries to estimate the likelihood-to-evidence ratio in the lower-dimensional space to reduce the complexity and increase the computational efficiency of the problem:
\begin{equation}
    \label{eqn:logR_compression}
    r_\phi (\theta,D) = f_\phi \left(s = C_\phi(x=D), \theta \right)
\end{equation}
$\texttt{PolySwyft}$ inherits this functionality via a neural compression network through $\texttt{swyft}$.

\subsubsection{Learning rate scheduling between rounds}
\label{subsec:lr_between_rounds}
When retraining a pre-trained neural network on a new concatenated dataset, one needs to adjust the initialisation of the learning rate to continue training to balance exploration and exploitation of the loss field. In practice, to accelerate compression, it's better to start with a high initial learning rate in the first rounds and then re-adjusting to lower rates when stabilised posterior estimates are found. We note that fine-tuning the learning rate scheduling between rounds is largely empirically driven, similar to the learning rate scheduling for an individual neural network round. However, there are indicators that the current scheduling might not be efficient, e.g. a flattening low compression in the earlier stages of the retraining due to low learning rates or sudden divergence of the compression as the learning rates are set to high. Observing the KL metric and readjusting subsequent retraining accordingly is a good practice.  
\subsubsection{Dynamic nested sampling}
\label{subsec:dynamic_live}
\begin{figure}
    \centering
    \includegraphics{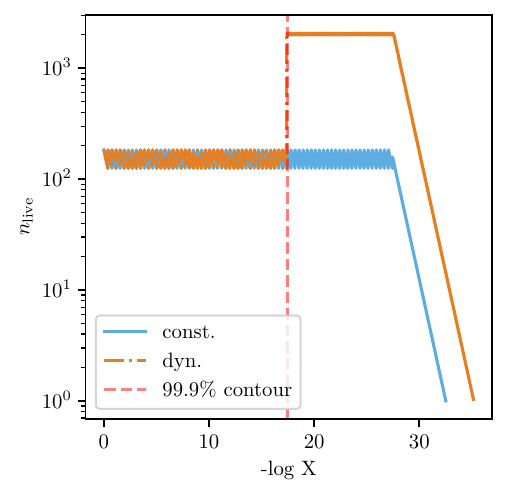}
    \caption{A simple dynamic nested sampling mechanism that increases the number of live points $n_{\mathrm{live}}$ at the $99.9\%$ posterior contour (red) that was found using a quick initial run (blue). The x-axis is in negative $\log X$ prior volume contraction scale. Here, the initial blue run determines the posterior contours of the current ratio estimator that a second run leverages for dynamically adjusting the live points at a given contour. In principle, the live point profile can be adjusted to any profile.}
    \label{fig:livepoints_boost}
\end{figure}

By adjusting the number of live points for a nested sampling, known as dynamic nested sampling \citep{higson_dynamic_2019}, one influences the distribution of the training dataset that is sequentially concatenated. First, we can include more samples from the relevant posterior region to the training dataset by dynamically increasing the number of live points at a given posterior region $\alpha$ contained within a specific $\log r$ contour. With $\texttt{PolySwyft}$, one can quickly introduce this boosting mechanism by executing two $\texttt{PolyChord}$ (or any dynamic nested sampler) runs in sequence. In the first iteration, we will execute $\texttt{PolyChord}$ with a constant low number of live points, here in the default settings $n_{\mathrm{live}} = 25 n_{\mathrm{dim}}$ to find the posterior regions of the NRE. In the second iteration, we use $\texttt{anesthetic}$ \citep{handley_anesthetic_2019} to find an $\alpha \%$ posterior region with the associated $\log r_{\alpha}$ contour that we can feed into $\texttt{PolyChord}$ to dynamically boost $n_{\mathrm{live}}$ at that contour line. Figure~\ref{fig:livepoints_boost} visualises this boosting functionality.

With this methodology, one actively steers the neural network's continual training towards the relevant posterior regions. One can also accelerate convergence, as the statistical expression introduced through each dynamic nested sampling-generated training dataset is usually higher than a standard run that contains a larger number of prior samples. This functionality is disabled by default.
We note that the dynamic nested sampling profile presented in Figure~\ref{fig:livepoints_boost} is the first approach to influence the distribution of the training dataset. Hence, we encourage probing a self-tuning dynamic live point profile for further research to maximise the statistical power introduced by the current dead points and increase sample efficiency.

\subsubsection{Noise resampling}
\label{subsec:noise_resamp}
Similarly, to accelerate the training of the neural network, we can introduce noise resampling as a method to increase the number of relevant training samples, where we can resample a set of $N$ jointly drawn samples $\{(D_{n,i}, \theta_i)\}_{n=1}^{N}$ for a given dead point $\theta_i$. Hence, noise resampling effectively resamples from the distribution $D \sim p(D|\theta)$ by adding noise around the dataset, saving on simulation costs. This functionality is disabled by default.

\subsubsection{Dead points post-processing}
\label{subsec:dead_trunc}
A classifier can encounter issues when too many samples are irrelevant to the ratio estimation problem, e.g., a significant portion of samples outside the posterior or vice versa. To mitigate this, we can use customizable post-processing of dead points after a nested sampling run in each round fed in as an optional function call. \texttt{PolySwyft} does not activate this functionality by default, instead using an unaltered set of dead points for retraining.

\section{Toy Problems}
\label{sec:Toy}
For each toy problem, we use 60 CPU cores on the Cambridge Service for Data-Driven Discovery (\href{https://www.csd3.cam.ac.uk/}{CSD3}) HPC cluster at the University of Cambridge. We use  \href{https://github.com/Lightning-AI/pytorch-lightning} {\texttt{pytorch-lightning}}'s distributed data-parallelism (DDP) methodology to accelerate neural network training. 

We use the default neural network structure of the $\texttt{swyft}$ package: A neural network that has a linear input layer and two residual blocks consisting of a series of batch normalisation, ReLU activation, linear layer and a dropout layer. For \texttt{swyft}'s truncation scheme, we use the default likelihood-to-evidence ratio selection threshold of $\epsilon_{\texttt{swyft}}  = 10^{-6}$ per truncation round.

For our neural network training schedule, we set the learning rate scheduling between rounds as  $\alpha_{\mathrm{init.}}(\mathrm{rd}) = \alpha_{\mathrm{init.}}(0) *\alpha_{\mathrm{decay}}^{n_{\mathrm{patience}} * \mathrm{rd}}$ with $\alpha_{\mathrm{init.}} (0)=0.001$. Within each round, we also set an exponential decaying learning rate schedule. The early stopping patience parameter is set to $n_{\mathrm{patience}} = 20$, while the decay rate $\alpha_{\mathrm{decay}}$ is varied with $0.01$ increments of $\alpha_{\mathrm{decay}} \in [0.95, 1.00]$ and choose the outcome that yield the best compression. We use the Adam optimiser \citep{kingma_adam_2017}, dropout probability of $p_{\mathrm{drop.}} = 0.3$ and a batch size of $n_{\mathrm{batch}} = 64$ throughout all experiments.
We also use no noise resampling, no dynamic nested sampling, no dead points post-processing and no data compression network.

\subsection{Multivariate linear Gaussian model (MVG)}
\label{subsec:MVG_example}

We define a linear Gaussian model $D = m + M*\theta \pm \sqrt{C}$, with a Gaussian prior on the parameters $\theta$, a Gaussian likelihood $p(D|\theta)$ with a model matrix $M$. We can then derive the posterior expression as: \begin{itemize}
    \item Prior: $\pi(\theta) \equiv \mathcal{N}(\mu, \Sigma)$
    \item Likelihood: $p(D|\theta) \equiv \mathcal{N}(m+M \theta, C)$
    \item Evidence: $p(D) \equiv  \mathcal{N}(m + M \mu, C + M \Sigma M^T)$
    \item Posterior: $p(\theta |D) = \mathcal{N}\left( \tilde M, \tilde C \right)$
\end{itemize}
with $\tilde M = (\Sigma^{-1} + M^T C^{-1} C)^{-1} (\Sigma^{-1} \mu + M^T C^{-1} (D-m))$ and $\tilde C = (\Sigma^{-1} + M^T C^{-1} M)^{-1}$. $M$ is a transformation matrix of $\dim(M) = (N_D, N_\theta)$. 
We use the Python package \texttt{lsbi}\footnote[1]{https://github.com/handley-lab/lsbi} to implement this MVG simulator and initialise the simulator with:
$\dim(\theta) = 5$, $\dim(D) = 100$, $\mu = 0$, a random Gaussian sample~$m$, a random Gaussian sample~$M$, $C = \mathbbm{1}$, and $\Sigma =  100 \times \mathbbm{1}$. We use this toy problem as an example of a relatively wide prior problem that is uncorrelated and independent.

We train the first NRE estimate with $N^{(0)} = 10.000$ prior samples to achieve a reasonable first estimate. We set the number of live points to $n_{\mathrm{live}} = 100 n_{\dim}$, generating $n_{\mathrm{dead}} \approx 15.000$ new dead points with each round.

\begin{figure*}
    \centering
    \includegraphics{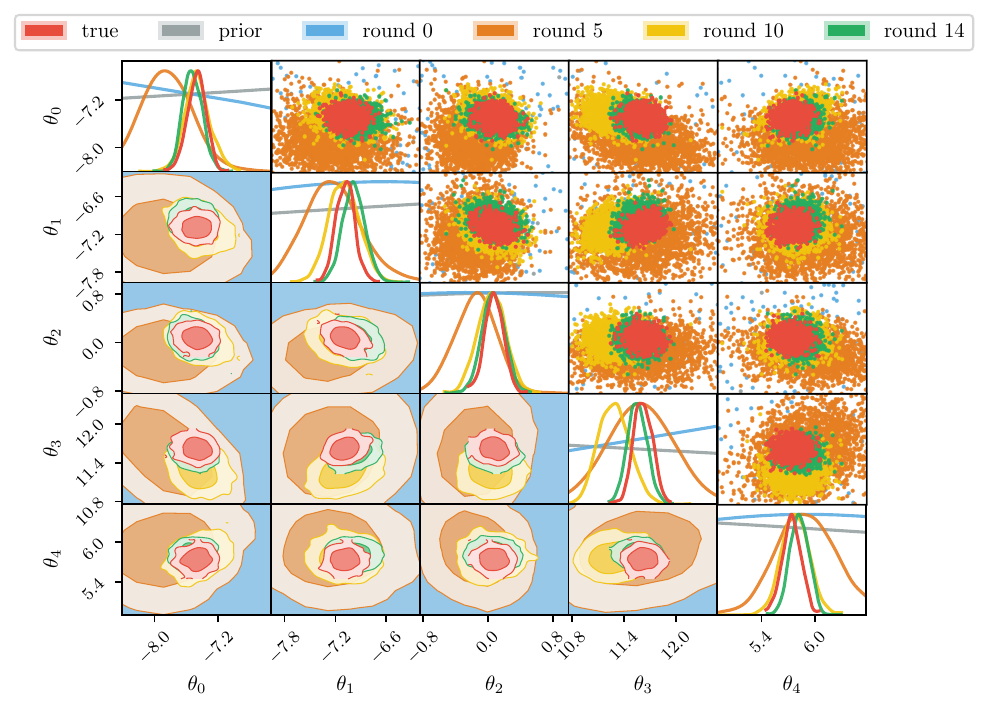}
    \caption{Estimated posterior distributions for a 5-dimensional MVG problem in a 100-dimensional data space. The posterior estimates become incrementally more accurate to the ground truth (red). The full prior distribution is shown in grey, overlaid by the posterior estimates. The kernel density estimates show $68\%$ (darker shade) and $95\%$ (brighter shade) iso-contour lines.}
    \label{fig:posterior_MVG}
\end{figure*}

\begin{figure}
    \centering
    \includegraphics{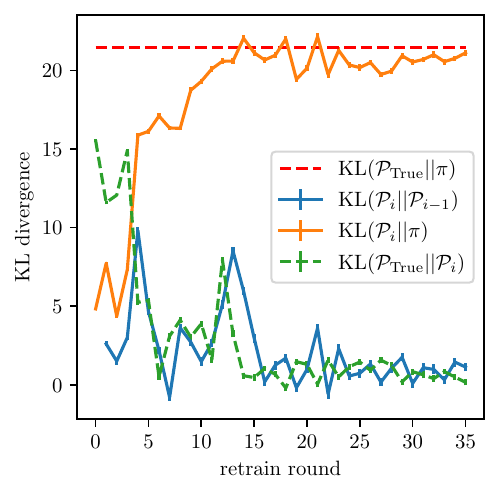}
    \caption{KL divergence for the MVG example. The red dashed line shows the analytical ground truth compression between the prior and the posterior. The blue line shows the comparison between the current and last posterior estimate. The orange line shows the compression from the prior to the posterior estimate. The green dashed line compares the ground truth and the current posterior estimate. In a real-world experiment, the dashed lines are not available.}
    \label{fig:MVG_KLdiv}
\end{figure}

With this setup, we show the KL-divergence plots and estimated posterior distributions in Figures~\ref{fig:posterior_MVG} and ~\ref{fig:MVG_KLdiv}. We present the posterior results through a ``triangle plot'' - generated with \texttt{anesthetic} - containing a 2-dimensional marginalised posterior Kernel Density Estimation (KDE) (lower triangle), but also the raw sample scatter plots (upper triangle, mirrored towards the diagonal) that generated the KDE. On the diagonal axis, the 1-d marginalised posterior KDEs are shown. 

As we have the analytical solution for this problem, we can also plot the $\mathrm{KL}(\mathcal{P}_{\mathrm{true}}||\mathcal{P}_{i})$ between the ground truth and the \texttt{PolySwyft} estimate. We reach convergence after 14 retraining rounds to the ground truth. \texttt{PolySwyft} needed $\sim \mathcal{O}(10^5)$ samples to find a posterior estimate for a given random observation. The CPU wallclock time of \texttt{PolySwyft} was $t \sim 1h$.

\begin{figure*}
    \centering
\includegraphics{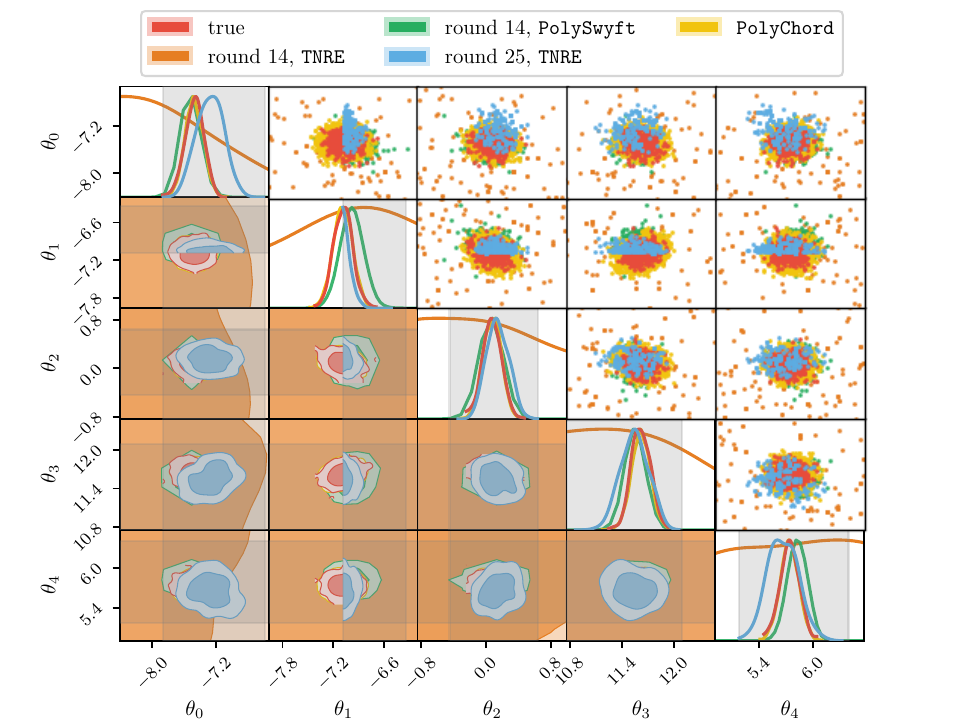}
    \caption{Comparison of \texttt{PolySwyft} and TNRE with its estimated posterior distributions for a 5-dimensional MVG problem in a 100-dimensional data space. The convergence at round 14 for \texttt{PolySwyft} (green) is comparable to the estimates at round 25 (blue) when using the prior truncation scheme (light grey box) of the TNRE. We note that TNRE's (automated) truncation scheme cuts off significant true posterior mass in $\theta_1$. $\texttt{PolyChord}$'s likelihood-based estimates are shown in yellow when using a standard initialisation scheme. }
    \label{fig:MVG_swyftVSpolyswyft}
\end{figure*}

As multivariate (dependent) priors are generally not supported for $\texttt{swyft}$'s rectangular truncation scheme, one has to decompose the multivariate Gaussian prior distribution with a diagonal covariance matrix $\Sigma$ into its marginalised independent 1-d components. For our toy example the prior initialisation changes to $\mu = 0 \rightarrow \mu_j = 0$ and $\Sigma = 100 \times \mathbbm{1} \rightarrow \sigma^2_j = 100$ for each marginal component $j$ to make it compatible for $\texttt{swyft}$'s truncation mechanism. We use $N^{(i)}=15.000$ samples in each round, sampled from the truncated region of the prior to train a new neural network. At round 14, $\texttt{PolySwyft}$ needed $\sum_{i=0}^{13} N^{(i)} \approx 200.000$ samples for convergence, hence, we use the TNRE's round 14 estimate for comparison. The TNRE posterior estimate can be seen in Figure~\ref{fig:MVG_swyftVSpolyswyft}. The posterior estimates of the TNRE at round 14 are significantly wider than the converged estimates by \texttt{PolySwyft} in the same round. Only when truncating further up until to round 25, the TNRE estimates resemble the ground truth. However, we note that the (automated) truncation bounds in the parameter $\theta_1$ constructed a wrong lower bound such that half of the distribution is excluded relative to the ground truth. For this example, $\texttt{PolySwyft}$ converged faster with approximately half the samples than the TNRE. Moreover, $\texttt{PolySwyft}$ does not suffer from issues regarding wrongly estimated truncation bound limits as it always does inference with the full prior.

We also include a standard $\texttt{PolyChord}$ run by using the known analytical likelihood. We initialise $\texttt{PolyChord}$ with the identical nested sampling initialisation of $\texttt{PolySwyft}$ with $n_{\mathrm{live}}=100 n_{\dim}$ and a termination criterion of $\epsilon_{\mathrm{NS}} < 0.001$ for a vanilla comparison of posterior estimates. For this MVG example, $\texttt{PolyChord}$ needed 1.8m likelihood calls, making $\texttt{PolySwyft}$ $90\%$ more efficient.

\subsection{Multivariate linear Gaussian mixture model (GMM)}
\label{subsec:GMM_example}
Similarly to the previous example, we can define a linear Gaussian mixture model:
\begin{itemize}
    \item Component Prior: $\pi(A) \equiv \mathrm{categorical}(\exp(\log w(A))$
    \item Prior: $\pi(\theta|A) \equiv \mathcal{N}(\mu, \Sigma)$
    \item Likelihood: $p(D|\theta,A) \equiv \mathcal{N}(m+M \theta, C)$
    \item Evidence: $p(D|A) \equiv  \mathcal{N}(m + M \mu, C + M \Sigma M' )$
    \item Posterior: $p(\theta |D,A) = \mathcal{N}\left(\mu + S M'C^{-1}(D - m - M \mu), S \right)$
\end{itemize}
with $S = (\Sigma^{-1} + M'C^{-1}M)^{-1}$. We use the Python package \texttt{lsbi} to implement this simulator, the same package as the MVG example, and initialise the simulator with: $\dim(\theta) = 5$, $\dim(D) = 100$, $\dim(A)=4$, random component weights $A$, a random Gaussian $\mu(A)$, a random Gaussian $m(A)$, a random Gaussian $M(A)$, $C = 4 \times \mathbbm{1}$, and $\Sigma(A)$ are random realisations of a Wishart distribution \citep{wishart_generalised_1928} to introduce correlated mixture components. With these initialisations, we generate a multimodal posterior example.

\begin{figure*}
\centering

\includegraphics{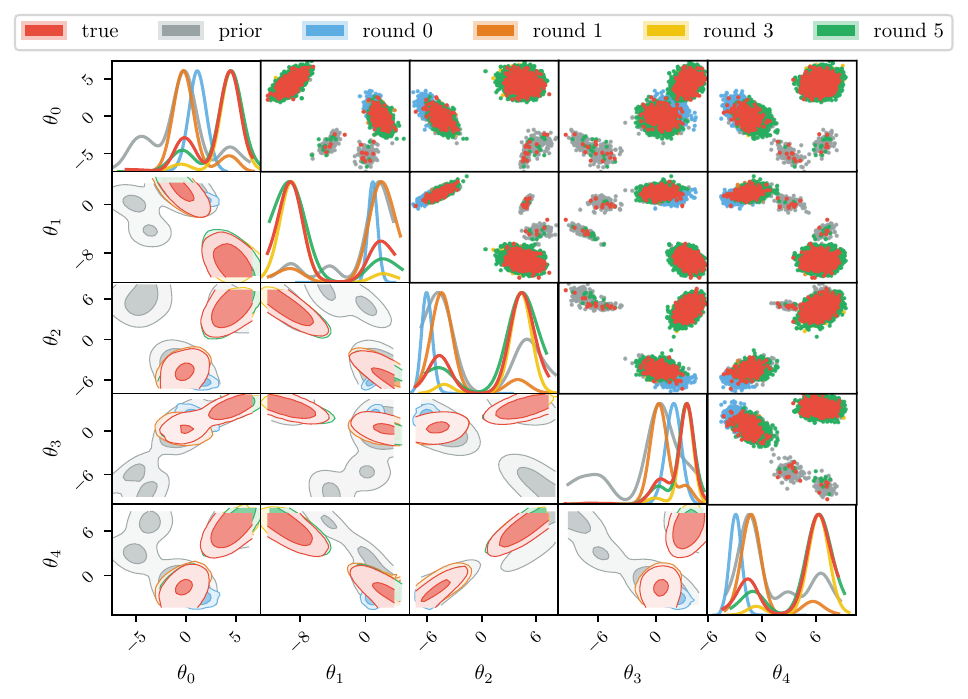}
    \caption{Estimated posterior distributions for a 5-dimensional GMM problem in a 100-dimensional data space. The posterior estimates become incrementally more accurate to the ground truth (red). The lower triangles are density estimates of 2-d marginalised parameter pairs with $68\%$ (dark shade) and $95\%$ (lighter shade) iso-contour lines. The diagonals are fully marginalised 1-d estimates, and the upper triangles are raw mirrored samples (towards diagonals) used to generate the density estimates.}

\label{fig:posterior_GMM}
\end{figure*}

\begin{figure*}
    \centering
    \includegraphics{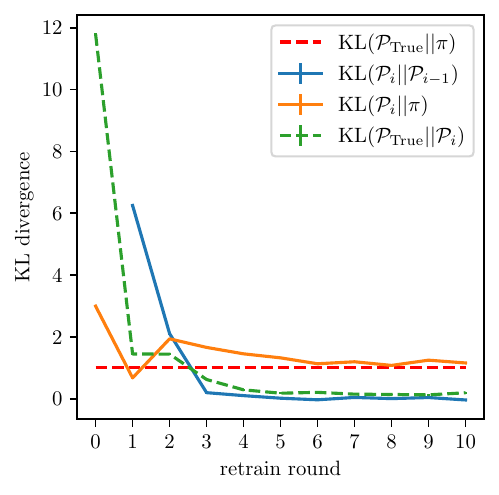}
    \caption{KL divergence for the GMM example. The dashed red line shows the analytical ground truth compression between the prior and the posterior. The blue line shows the comparison between the current and the last posterior estimate. The orange line shows the compression from the prior to the posterior estimate. The green dashed line compares the ground truth and the current posterior estimate.}
    \label{fig:GMM_KLdiv}
\end{figure*}

For this toy problem, we initially train an NRE on $N^{(0)}=30.000$ examples and set a constant set of live points $n_{\mathrm{live}} = 500 n_{\dim}$ as we need more live point density to detect and explore complex multimodal posterior distributions. With these settings, we add around $n_{\mathrm{dead}} \approx 25.000$ dead points to the training dataset for each round. \texttt{PolySwyft} converges after 5 rounds to the ground truth needing $\mathcal{O}(10^5)$ samples. The CPU wallclock time of \texttt{PolySwyft} was $t \sim 2h$.
We present the estimated posterior distributions in Figure~\ref{fig:posterior_GMM} and the KL divergence plots in Figure~\ref{fig:GMM_KLdiv}. 
One notices that in the initial round 0, the estimated posterior is unimodal, which is seen as a larger compression $\mathrm{KL}(P_0 ||\pi$) than its analytical answer $\mathrm{KL}(P_{\mathrm{true}} ||\pi$) that is multimodal. However, due to the increased live point density, one can correct the initial posterior estimate to the ground truth as we still explore the full prior space in the subsequent nested sampling run. Hence, geometrical truncation schemes of the prior based on the estimates of round 0 could fail to detect the multimodality, as these schemes can construct new truncation bounds that only enclose one of the modes, therefore enforcing a unimodal posterior for subsequent training efforts.

\begin{figure*}
\centering
\includegraphics{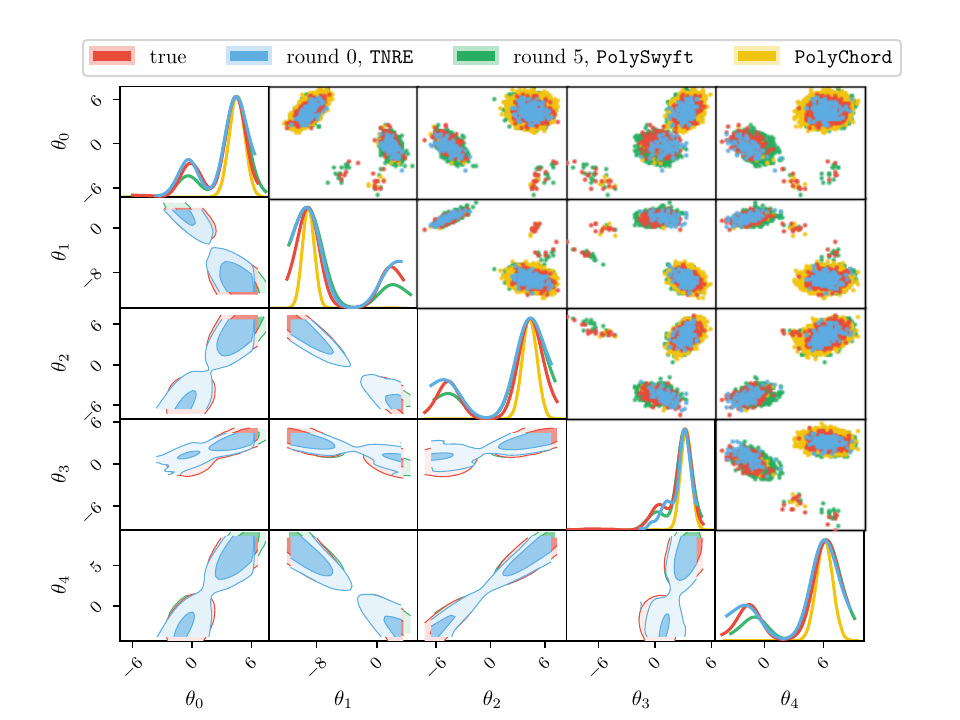}
    \caption{Estimated posterior distributions for a 5-dimensional GMM problem in a 100-dimensional data space. The posterior estimates of the TNRE (light blue), \texttt{PolySwyft} (green), \texttt{PolyChord} (yellow) and the ground truth (red) are compared. Here, the TNRE was trained on the cumulative number of samples in round 0 that \texttt{PolySwyft} needed to reach convergence. For \texttt{PolyChord}, we use the standard settings of $n_{\mathrm{live}}=500 n_{\dim}$ and $\epsilon_{\mathrm{NS}}<0.001$. The lower triangles are density estimates with $68\%$ and $95\%$ iso-contour lines of 2-d marginalised parameter pairs, the diagonals are fully marginalised 1-d estimates, and the upper triangles are raw mirrored samples towards the diagonal used to generate the density estimates. We note the tendency of the density estimates to smooth out the minor clusters in all three estimates. However, the raw samples show that the TNRE does not capture the minor clusters while $\texttt{PolySwyft}$ and \texttt{PolyChord} do.}

\label{fig:posterior_GMM_swyft_comparison}
\end{figure*}

For this example, we can not use the truncation scheme of the TNRE, as the rectangular truncation scheme requires independent and uncorrelated priors with known cumulative distribution functions (CDFs) to construct new truncation bounds and the inverse CDF for the inverse transformation sampling \citep{devroye_non-uniform_1986} procedure to sample within the bounds. For our GMM problem, with categorical component priors and correlated component covariances, an analytical expression for the CDF and inverse CDF is not known. One can approximate these quantities through numerical efforts, however, this requires significant feature expansion on the TNRE's codebase to support arbitrarily complex distributions. Whereas \texttt{PolySwyft} can easily be extended to any (correlated and dependent) prior distribution with a known hypercube sampling procedure. For the GMM problem, this prior sampling procedure is implemented by \texttt{lsbi} using bijectors; however, we note that bijectors can introduce their own set of challenges.

Hence, for our comparison analysis we train the TNRE on the cumulative number of samples that $\texttt{PolySwyft}$ needed in 5 rounds to reach convergence, here $\sum_{i=0}^{4}N^{(i)} \approx 160.000$. We present the results of the TNRE posterior estimates in Figure~\ref{fig:posterior_GMM_swyft_comparison}.
We notice that the TNRE's posterior is a good estimate of the ground truth. However, in the TNRE estimates, one of the minor clusters next to the two main clusters was not approximated, while $\texttt{PolySwyft}$ could detect them. A subsequent geometric truncation scheme based on this TNRE estimate could likely fail and exclude this multimodality. This implies that for arbitrarily multimodal distributions, the initial estimate of TNRE has to be trained on a dataset that detects all the clusters, thus solving the full problem in the first training effort ``round 0'', while $\texttt{PolySwyft}$ does not require this. $\texttt{PolySwyft}$ can correct for initial bad estimates, such as not detecting all of the modes, by always sampling from the full prior and exploring the likelihood with its current set of live points. This self-correction mechanism is evident in Figure~\ref{fig:posterior_GMM}, in which $\texttt{PolySwyft}$ started as a unimodal estimate in round 0. We note that in practice, this self-correction mechanism depends on the density of the live points; therefore, $\texttt{PolySwyft}$ can also fail to detect all modes if the density of the live points is not sufficient. 

This live point population can be troublesome when combined with a prior that uses bijectors to sample from the hypercube space. In the \texttt{PolyChord} posterior estimates shown in Figure~\ref{fig:posterior_GMM_swyft_comparison}, the kernel density estimations of the 1-d and 2-d marginals struggle to capture the second largest cluster. This is due to the geometric distortion introduced by the reparametrisation of bijectors, which distorts the likelihood contours through the bijective mapping between a hypercube and a Gaussian mixture model \citep{betancourt_conceptual_2018}. Moreover, this distortion bias does not occur with \texttt{PolySwyft} and the TNRE that skipped the bijectors by directly sampling from the original prior and using the same likelihood function that \texttt{PolyChord} used as a simulator. As an alternative hypercube sampling procedure is not provided by \texttt{lsbi}, which is required by \texttt{PolyChord}, we leave this sampling correction for future work, with e.g.\ \cite{yallup2025nested}'s nested sampler allowing sampling from any prior distribution. Overall, the scatter plot reveals that each cluster was detected by \texttt{PolyChord} (although overlaid by the other samples), but with lower component weights for the second major cluster. \texttt{PolyChord} needed 1.5m likelihood calls to recover this posterior estimate, a $90\%$ efficiency difference to $\texttt{PolySwyft}$.

\subsection{Cosmic Microwave Background (CMB) example}
The third example we apply \texttt{PolySwyft} on is the recovery of the posterior distribution of the cosmological parameters using Planck data of the cosmic microwave background (CMB) ~\citep{aghanim_planck_2020, planck_collaboration_planck_2020}. As we assume no likelihood within the SBI paradigm, we will need a CMB simulator to generate the spectra given some prior distribution. For this task, we use the CMB emulator library \texttt{cmb-likelihood}\footnote[1]{https://github.com/htjb/cmb-likelihood} that provides an interface for \texttt{CAMB} ~\citep{lewis_camb_2011} and \texttt{CosmoPower} ~\citep{mancini_cosmopower_2022}. We chose \texttt{CosmoPower}, a neural network-driven emulator, as the CMB simulator due to its efficiency of spectra generation over \texttt{CAMB} to mitigate the simulation time bottleneck of SBI. 

As for the observation, we use a simulated observation using \texttt{CosmoPower} conditioned on the best-fit parameters of the Planck study: the baryon density $\Omega_b h^2 = 0.022$, dark matter density $\Omega_c h^2 = 0.120$, optical depth $\tau = 0.055$, scalar spectral index $n_s = 0.965$, marginalised power spectrum amplitude $\ln 10^{10}A_s = 3.0$, and the reduced Hubble constant $h =0.67$. We conduct the following binning to maintain binning parity across simulations: $\Delta l = 1, \; \forall l \in [2,30]$, $\Delta l = 30, \; \forall l \in [30,2508]$ and the last bin containing the remainder. With this binning procedure, the resulting dimensionality of the problem reduces to $\dim(\theta,D)=(6,111)$. 

After the binning procedure, we pre-process the CMB power spectra for the neural network input layer of the NRE through a series of normalisation, log-transformation and standardisation, which has similarly been applied to train a Variational Auto-Encoder (VAE) with \texttt{CosmoPower} \citep{piras_mathrmensuremathlambdamathrmcdm_2025}.
For the normalisation step, we use the simulated observation as the reference spectrum. For the standardisation, we use the initial prior-sampled spectra to estimate the normalised logarithmic mean and standard deviation.

We set uniform priors on the CMB parameters baryon density $\omega_{\mathrm{b}}$, dark matter density $\omega_{\mathrm{CMB}}$, optical depth $\tau_{\mathrm{reio}}$, scalar spectral index $n_{\mathrm{s}}$, initial super-horizon amplitude of curvature perturbations $\ln 10^{10} A_s$ and Hubble parameter $h$ to generate simulations with \texttt{CosmoPower}. The prior ranges are shown in Table~\ref{tab:CMB_priors} and are slightly tighter boundaries than those presented in \texttt{CosmoPower} as an emulator trained on extreme cosmological parameter combinations of a physics-driven simulator can fail.
\begin{table}
    \centering
    \begin{tabular}{llll}
\toprule
{}
Parameter & $\theta_{\min}$ & $\theta_{\max}$ & type \\
\midrule
          $\omega_{\mathrm{b}}$ &  0.005 & 0.04 & uniform \\
         $\omega_{\mathrm{CMB}}$ &  0.08 & 0.21 & uniform \\
         $\tau_{\mathrm{reio}}$ &  0.01 & 0.16 & uniform \\
         $n_{\mathrm{s}}$ &  0.8 & 1.2 & uniform \\
         $\ln 10^{10} A_s$ &  2.6 & 3.8 & uniform \\
         $h$ &  0.5 & 0.9 & uniform \\
\bottomrule
    \end{tabular}
    \caption{Uniform prior ranges for the CMB parameters on \texttt{CosmoPower}.}
    \label{tab:CMB_priors}
\end{table}
With \texttt{PolySwyft}, we use a standard nested sampling run with $n_{\mathrm{live}} = 100 n_{\dim}$ and apply no additional noise resampling, no dead point post-processing, no dynamic nested sampling, no data compression network and terminate the nested sampling run with the standard evidence termination criterion of \texttt{PolyChord}.

\begin{figure*}
    \centering
    \includegraphics{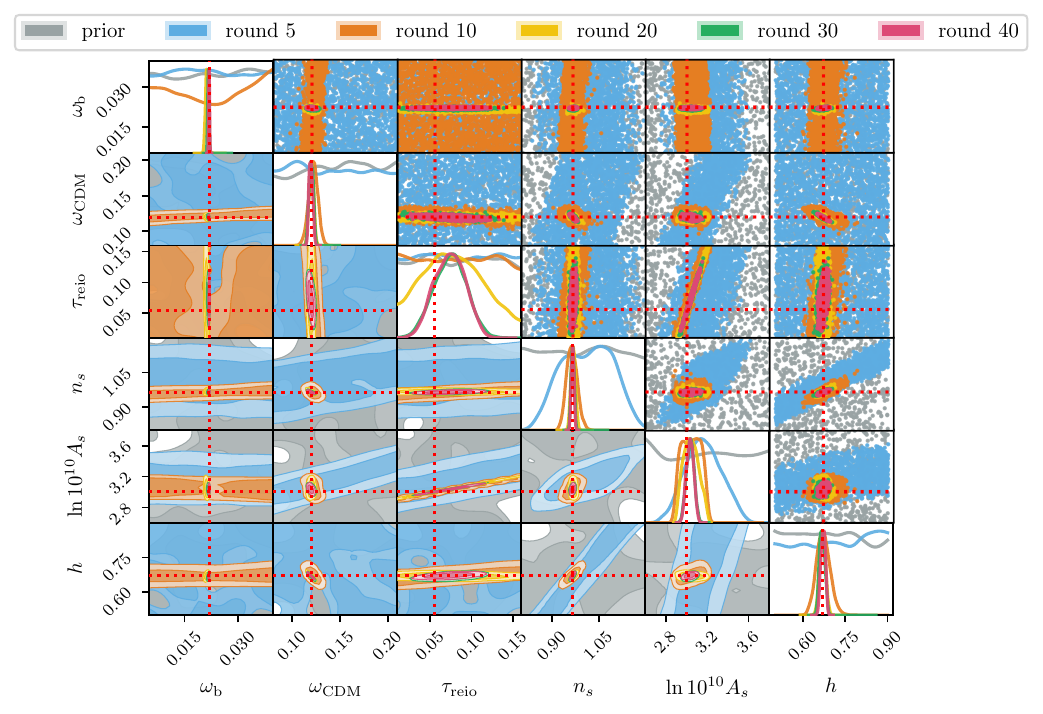}
    \caption{Estimated posterior with \texttt{PolySwyft} for a CMB example using Planck data and CosmoPower as a simulator. The red dashed lines are the fiducial estimates reported by Planck: $\hat{\omega}_{\mathrm{b}} = 0.022$, $\hat {\omega}_{\mathrm{CMB}}=0.12$, $\hat {\tau}_{\mathrm{reio}}=0.055$, $\hat {n}_{\mathrm{s}}=0.965$,  $\ln 10^{10} \hat{A}_s=3.0$,  $\hat h =0.67$.}
    \label{fig:posterior_CMB}
\end{figure*}

\begin{figure}
    \centering
\includegraphics{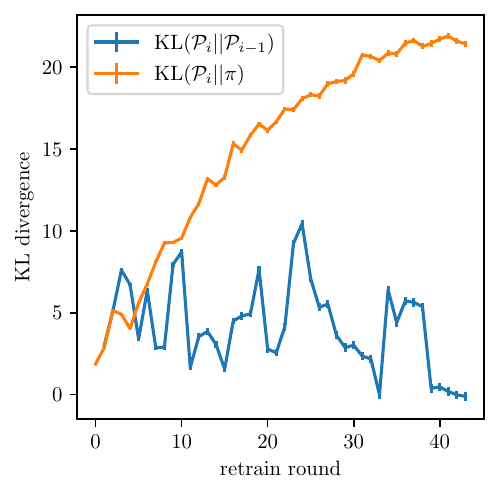}
    \caption{KL divergence for the CMB example. The blue line shows the comparison between the current and last posterior estimate. The orange curve shows the compression from the prior to the posterior estimate.}
    \label{fig:CMB_KLdiv}
\end{figure}

With these settings, we add about $n_{\mathrm{dead}} \approx 15.000$ new samples per round to the initial prior samples dataset of size $N^{(0)}=10.000$.  The resulting posterior estimates and KL convergence diagnostics are shown in Figures~\ref{fig:posterior_CMB} and~\ref{fig:CMB_KLdiv}. After 40 retraining rounds, we recover the best-fit estimates of the cosmological parameters of the Planck collaboration within the $2\sigma$ regions.  In the termination round, \texttt{PolySwyft} needed $N \approx 650.000$ samples to reach convergence at round 40. The CPU wallclock time of \texttt{PolySwyft} was $t \sim 2h$.

\begin{figure*}
    \centering
\includegraphics{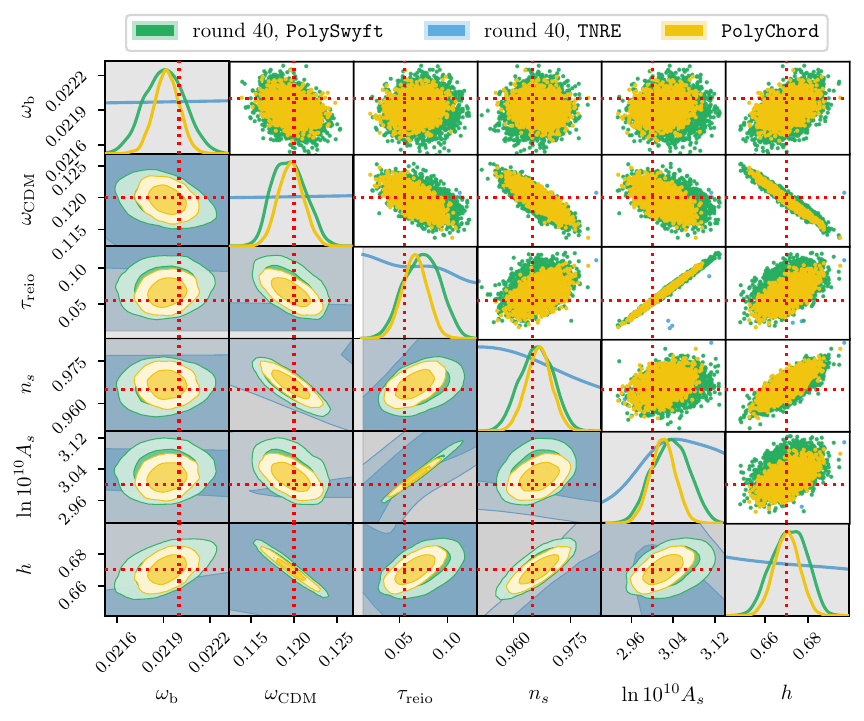}
    \caption{CMB posterior comparison of $\texttt{PolySwyft}$ (green), TNRE (blue) with its prior truncation bound (grey) and likelihood-based $\texttt{PolyChord}$ (yellow). While the TNRE estimates could not significantly constrain the fiducial estimates (red dashed), $\texttt{PolySwyft}$ and $\texttt{PolyChord}$'s estimates show similar recovery of the fiducial estimates.}
    \label{fig:CMB_swyftVSpolyswyft}
\end{figure*}

\begin{figure*}
    \centering
\includegraphics{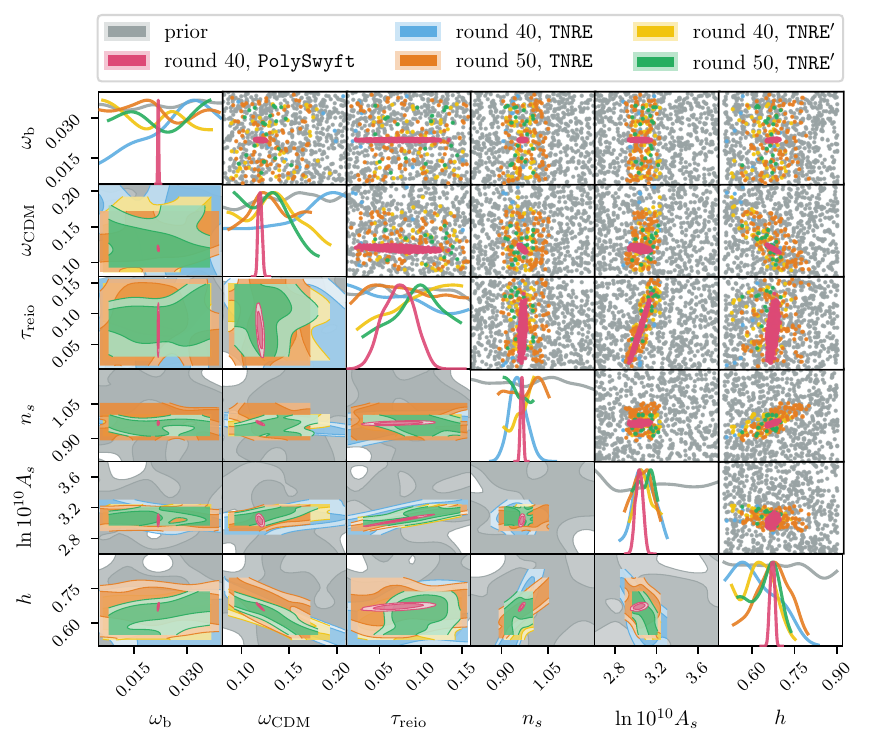}
    \caption{CMB posterior comparison of $\texttt{PolySwyft}$ (red) with a TNRE (blue and orange),
    and a more aggressively truncated TNRE' (yellow and green) with $\epsilon_{\mathrm
    {swyft}}' = 10^3 \times \epsilon_{\mathrm
    {swyft}}$. After 40 truncation rounds, we note that both the TNRE truncation schemes do not significantly constrain $\omega_b$, $\omega_{\mathrm{CDM}}$ and $\tau_{\mathrm{reio}}$. While $n_s$, $\ln 10^{10}A_s$ and $h$ show signs of compression; however, they have significantly higher variance compared to the $\texttt{PolySwyft}$ estimates. Moreover, 10 further rounds of truncation in both default and aggressive truncation schemes do not yield satisfactory results, indicating an overall struggle for the TNRE for this CMB problem.}
    \label{fig:CMB_swyftVSpolyswyft_extended}
\end{figure*}

As the priors on the cosmological parameters are independent, one does not need additional refactoring to use TNRE's truncation scheme. When comparing these posterior estimates with the TNRE in Figures~\ref{fig:CMB_swyftVSpolyswyft} and~\ref{fig:CMB_swyftVSpolyswyft_extended}, we see that the TNRE estimates are less constrained in every parameter, suggesting the TNRE struggles to recover the parameters for this problem. This is further confirmed when the truncation is continued over 40 rounds, and no significant progress is achieved with the TNRE. Moreover, we also rerun the truncation by setting up a more aggressive truncation scheme of $\epsilon_{\mathrm{swyft}}' = 10^{-3} \equiv  10^3 \times \epsilon_{\texttt{swyft}}$. However, this did not significantly impact the posterior recovery. Overall, for this CMB problem, the TNRE struggles to recover the Planck estimates.

To compare $\texttt{PolySwyft}$'s estimates with $\texttt{PolyChord}$, we use \texttt{CosmoPower} to construct a likelihood through \texttt{cmb-likelihood}. We note that this likelihood is a generative, foreground-free, full sky, cosmic-variance-limited likelihood, implementing the underlying Wishart distribution. This is highly idealised compared to real data, but gives a good estimate of what such estimates would have on a full generative Planck likelihood, the latter of which would be easier to sample due to the reduction in information content from foregrounds and sky cuts. With this likelihood function, we use a vanilla $\texttt{PolyChord}$ run, and compare the nested sampling estimate with $\texttt{PolySwyft}$ in Figure~\ref{fig:CMB_swyftVSpolyswyft}. We note that $\texttt{PolySwyft}$'s posterior estimate matches the \texttt{PolyChord} estimate, with a slight tendency of higher variance, which is due to the difference of using an exact likelihood-expression versus an approximation found via simulations.

For this CMB example, $\texttt{PolyChord}$ needed 2.4 million likelihood calls, while $\texttt{PolySwyft}$ converged with 650 thousand simulator calls, a $\sim 75\%$ difference in efficiency.

\section{Future improvements of \texttt{PolySwyft}}
\label{sec:future_improv}
We presented various methodologies to improve upon the retraining cycle of \texttt{PolySwyft} in section~\ref{subsec:optimis}. To summarise, one can alter the retraining dataset by changing the dead measure distribution via dynamic nested sampling or through post-processing operations using truncation/early termination, compression or noise resampling schemes. As these methods are still applied in an ad-hoc manner, further research efforts must be spent on unifying these methodologies into a common framework.

For instance, choosing a posterior contour level $\alpha$ to adjust the live points is empirically driven, as seen in the live points profile shown in Figure~\ref{fig:livepoints_boost}. Further research will need to investigate an automated/self-tuning decision criteria that use current KL-divergence metrics and the expected statistical power of the dead points relative to the current training set to decide on the dynamic nested sampling profile.

The toy examples presented did not need any training data optimisation and were only optimised on the network side, e.g., the learning rate scheduling. However, for more complex problems, the standard settings might fail, and the aforementioned additional optimisation methods can be used to accelerate and fine-tune training. 

Moreover, it is a subject of future work to assess how scalable \texttt{PolySwyft} is to higher dimensions, as we only tested for dimensionality $\dim(\theta,D) \sim (6,111)$.
Studies by \cite{alvey_what_2023, bhardwaj_sequential_2023, anaumontel_scalable_2024} have shown that significant scalability can be achieved by either introducing marginal estimation (MNRE) instead of joint density estimation (NRE) or using an autoregressive model (ANRE) by decomposing a joint density into its product of 1-d conditional densities.

Therefore, further research directions to increase scalability and/or sample efficiency can entail extending \texttt{PolySwyft}'s capabilities in three ways: \begin{itemize}
    \item Introducing new algorithmic components such as marginal (NS\textbf{M}NRE), autoregressive models (NS\textbf{A}NRE) or other methods.
    \item Optimising the dead measure by changing the termination criterion of nested sampling, introduction of dynamic nested sampling, dead points post-processing, noise resampling or other techniques.
    \item A mixture of both approaches.
\end{itemize}

\section{Conclusions}
\label{sec:Conclusion}
With \texttt{PolySwyft}, we address the limitations of TNREs when dealing with complex posteriors that can be multimodal. \texttt{swyft}, an implementation of T(M)NRE, iteratively truncates the prior of the parameter space by constructing shrinking supporting regions until a termination criterion is reached. This truncation scheme is known to be limiting for highly multimodal or complex-shaped problems, and its sampling inefficiency increases exponentially with the number of dimensions.

We, therefore, propose \texttt{PolySwyft} that merges nested sampling (NS) and neural ratio estimation (NRE) into a common framework (NSNRE) to explore multimodal and complex problems marginal-free, where the likelihood is analytically intractable. \texttt{PolySwyft}, an implementation of an NSNRE method, uses \texttt{PolyChord} and \texttt{swyft}, two algorithms that are commonly used in the respective methodological area. We show with the work that there are intrinsic synergies between nested sampling and simulation-based inference that can expand on the current limitations of truncation-based methods of T(M)NREs.

More concretely, \texttt{PolySwyft} defines a termination criterion using the KL divergence between NRE estimates $\mathrm{KL}(\mathcal{P}_i || \mathcal{P}_{i-1})$ and its compression relative to the prior $\mathrm{KL}(\mathcal{P}_i || \pi )$ to assess whether convergence is reached. 

We demonstrate \texttt{PolySwyft} on a multivariate linear Gaussian model of dimensionality $\dim( \theta,D)= (5, 100)$, a multivariate linear Gaussian mixture model with dimensionality $\dim(\theta,D)= (5, 100)$ and a cosmological CMB example with $\dim(\theta,D) = (6, 111)$. We show that \texttt{PolySwyft} achieves posterior estimates that recover the ground truth (if available) of the underlying analytical problem or the best-fit estimates (for the CMB example) and achieves convergence with fewer samples ($\sim 50\%$) than the TNRE and fewer samples ($>75\%$) defined by the number of likelihood calls than the nested sampler \texttt{PolyChord}.

Further research efforts are required to increase the sample efficiency and scalability of $\texttt{PolySwyft}$. For instance, $\texttt{swyft}$'s T\textbf{M}NRE, i.e.\ marginal estimation capabilities and the algorithms' recommended settings, are shown to be even more efficient ($>90 \%$) than traditional likelihood-based methods such as MCMC or nested sampling. Hence, potential efforts to optimise $\texttt{PolySwyft}$'s sample efficiency could entail introducing marginal estimation (NS\textbf{M}NRE), autoregressive models (NS\textbf{A}NRE), a different termination criterion of the nested sampling cycle, the introduction of dynamic nested sampling, dead point post-processing in each round, noise resampling or other novel techniques.

\section*{Acknowledgements}
KHS would like to thank the Hans Werthén Foundation, the Alan Turing Institute and G-Research for providing the necessary funding for developing this algorithm. WH would like to thank the Royal Society University Research Fellowship. C.W. received funding from the European Research Council (ERC) under the European Union’s Horizon 2020 research and innovation programme (Grant agreement No. 864035 – UnDark). EdLA is supported by an Ernest Rutherford fellowship.
This work was performed using resources provided by the Cambridge Service for Data Driven Discovery (CSD3) operated by the University of Cambridge Research Computing Service (www.csd3.cam.ac.uk), provided by Dell EMC and Intel using Tier-2 funding from the Engineering and Physical Sciences Research Council (capital grant EP/T022159/1), and DiRAC funding from the Science and Technology Facilities Council (www.dirac.ac.uk).

\section*{Data Availability}
The code of $\texttt{PolySwyft}$ is available on GitHub: https://github.com/kilian1103/PolySwyft



\bibliographystyle{mnras}

\bibliography{oja_template}

\end{document}